\documentclass[conference]{IEEEtran}
\IEEEoverridecommandlockouts
\usepackage{cite}
\usepackage{amsmath,amssymb,amsfonts}
\usepackage{algorithmic}
\usepackage{graphicx}
\usepackage{textcomp}
\usepackage{xcolor}
\usepackage{tikz}
\usepackage{etoolbox}
\usepackage{balance}

\DeclareRobustCommand*{\IEEEauthorrefmark}[1]{%
  \raisebox{0pt}[0pt][0pt]{\textsuperscript{\footnotesize #1}}%
}

\newcommand{\ballnumber}[1]{\tikz[baseline=(myanchor.base)] \node[circle,fill=.,inner sep=1pt] (myanchor) {\color{-.}\bfseries\footnotesize #1};}

\def\BibTeX{{\rm B\kern-.05em{\sc i\kern-.025em b}\kern-.08em
    T\kern-.1667em\lower.7ex\hbox{E}\kern-.125emX}}

\title{Single-Cell Universal Logic-in-Memory Using 2T-nC FeRAM: An Area and Energy-Efficient Approach for Bulk Bitwise Computation}

\author{
\IEEEauthorblockN{
Rudra Biswas\IEEEauthorrefmark{1},
Jiahui Duan\IEEEauthorrefmark{2},
Shan Deng\IEEEauthorrefmark{2},
Xuezhong Niu\IEEEauthorrefmark{2},
Yixin Qin\IEEEauthorrefmark{2},
Prapti Panigrahi\IEEEauthorrefmark{1},
Varun Parekh\IEEEauthorrefmark{1},\\
Rajiv Joshi\IEEEauthorrefmark{3},
Kai Ni\IEEEauthorrefmark{2},
and Vijaykrishnan Narayanan\IEEEauthorrefmark{1}
}
\IEEEauthorblockA{\IEEEauthorrefmark{1}The Pennsylvania State University, University Park, PA, USA\\
}
\IEEEauthorblockA{\IEEEauthorrefmark{2}University of Notre Dame, South Bend, IN, USA}
\IEEEauthorblockA{\IEEEauthorrefmark{3}IBM T. J. Watson Research Center, Yorktown Heights, NY, USA\\
Emails: rfb5659@psu.edu}
}

\begin{document}

\maketitle
\begin{abstract}
% The growing demands of data-intensive workloads challenge conventional computing architectures, where data movement between memory and processing units creates a significant energy and performance bottleneck. This work presents a comprehensive analytical study of 2T-nC ferroelectric RAM (FeRAM) as a non-volatile alternative for performing efficient logic-in-memory (LiM) operations. We demonstrate that the quasi-nondestructive readout (QNRO) mechanism intrinsic to 2T-nC FeRAM can be exploited to perform universal bitwise logic directly within the memory array. This approach supports NOT operations without cell modification and is extended to universal NAND/NOR logic via a MINORITY function, which we validate through SPICE simulations and experimental data. To assess its practical impact, we evaluate the design across eight real-world, data-intensive applications and validate its thermal feasibility for vertical 3D integration. We also experimental demonstrate the MINORITY function in the fabricated 2T-nC FeRAM cell. Compared to a conventional DRAM-based system, our 2T-nC FeRAM LiM approach achieves a 2× improvement in performance and a 2.5× reduction in energy consumption. This research establishes 2T-nC FeRAM as a viable and efficient technology for building next-generation computing systems, offering enhanced in-memory functionality while overcoming critical limitations of traditional memory hierarchies.
This work presents a novel approach to configure 2T-nC ferroelectric RAM (FeRAM) for performing single cell logic-in-memory operations, highlighting its advantages in energy-efficient computation over conventional DRAM-based approaches. Unlike conventional 1T-1C dynamic RAM (DRAM), which incurs refresh overhead, 2T-nC FeRAM offers a promising alternative as a non-volatile memory solution with low energy consumption. %%Our key findings include: i) The implementation of Bitwise logic operations in 1T-1C FeRAM for in-memory computation; ii) The unique quasi-nondestructive readout (QNRO) sensing mechanism in 2T-nC FeRAM inherently enables NOT logic without requiring cell modification, unlike 1T-1C FeRAM or DRAM; iii) Universal bitwise logic implementation in 2T-nC FeRAM leveraging QNRO sensing; iv) Using SPICE simulations and experimental validation, we demonstrate the implementation of the MAJORITY function in 1T-1C FeRAM, including associated AND/OR logic, as well as the MINORITY function in 2T-nC FeRAM, supporting universal NAND and NOR logic; v) Explain vertical 3D integration feasibility and advantages vi) Evaluate eight real-world, data-intensive applications that utilize bulk bitwise operations implemented via logic-in-memory. 
%Our key findings include the implementation of bitwise logic operations leveraging the unique quasi-nondestructive readout (QNRO) sensing mechanism in 2T-nC FeRAM inherently enables NOT logic without requiring cell modification, unlike 1T-1C DRAM. Furthermore, we demonstrate universal bitwise logic implementation in 2T-nC FeRAM leveraging QNRO sensing. Using SPICE simulations and experimental validation, we implement the MINORITY function in 2T-nC FeRAM. We also evaluate the feasibility and density advantages of vertical 3D integration and assess eight real-world, data-intensive applications utilizing bulk bitwise operations enabled by logic-in-memory. Our results show 2× higher performance and 2.5× lower energy consumption by 2T-nC FeRAM compared to DRAM. This research emphasizes the potential of 2T-nC FeRAM for logic-in-memory applications, offering both improved functionality and performance over traditional DRAM.
Our key findings include the potential of quasi-nondestructive readout (QNRO) sensing in 2T-nC FeRAM for logic-in-memory (LiM) applications, demonstrating its inherent capability to perform inverting logic without requiring external modifications, a feature absent in traditional 1T-1C DRAM. We successfully implement the MINORITY function within a single cell of 2T-nC FeRAM, enabling universal NAND and NOR logic, validated through SPICE simulations and experimental data. Additionally, the research investigates the feasibility of 3D integration with 2T-nC FeRAM, showing substantial improvements in storage and computational density, facilitating bulk-bitwise computation. Our evaluation of eight real-world, data-intensive applications reveals that 2T-nC FeRAM achieves 2× higher performance and 2.5× lower energy consumption compared to DRAM. Furthermore, the thermal stability of stacked 2T-nC FeRAM is validated, confirming its reliable operation when integrated on a compute die. These findings emphasize the advantages of 2T-nC FeRAM for LiM, offering superior performance and energy efficiency over conventional DRAM.

\end{abstract}


\section{Introduction}
Rapid growth of data-intensive applications has triggered the search for energy-efficient and high-performance computation. %Today’s computers are built on 
Conventional von Neumann architectures incur high latency and energy overhead due to frequent data movement between computational units and memories. To address this, Logic-in-Memory (LiM) has emerged as a promising paradigm, enabling computation within memory to minimize data transfers. LiM architectures have been extensively explored using both volatile %(e.g., SRAM \cite{8310397} and DRAM \cite{gao2019computedram}) 
and non-volatile %(e.g., RRAM, PCM \cite{article}, and ferroelectric devices \cite{yang2020memory, 9045150, 10900656})
memories \cite{gao2019computedram}, \cite{article}. %LiM architectures commonly employ volatile memories, such as SRAM \cite{8310397} and DRAM \cite{gao2019computedram}. 

Sheshadri et al. \cite{seshadri2017ambit} demonstrated fundamental logic operations in DRAM with AND, OR operations within 3 cells and NOT operations using external circuit modifications; however, such volatile memories have destructive sensing and suffer from low density, limiting storage and performance gains. Non-volatile ferroelectric memories (FeRAM) have gained interest due to their low-energy write operations enabled by external electric field-induced polarization switching~\cite{endoh2016overview,vlsid25}. The scalability and CMOS compatibility of ferroelectric HfO$_2$ have further revived interest in FeRAM-based LiM \cite{ramaswamy2023nvdram, okuno2023highly}. To leverage FeRAM's potential, Slesazeck et al. \cite{slesazeck20192tnc} proposed a 2T-nC FeRAM cell, suitable for compute-in-memory applications. However, challenges such as destructive read operations and write-back requirements remain.%within memory using DRAM technology, incorporating cell modifications to enhance functionality. However, the low memory density of these volatile memories limits both data storage capacity and performance improvements. 
%Significant research has focused on high-density non-volatile memories (NVMs) such as RRAM , PCM \cite{article}, and ferroelectric devices \cite{yang2020memory, 9045150, 10900656}. Among these, ferroelectric memory devices are particularly notable due to their low energy write operations enabled by external electric field induced polarization switching, resulting in an almost negligible static write current \cite{endoh2016overview}.Recent advantages in scalable CMOS-compatible ferroelectric HfO$_2$ have resurged the interest in ferroelectric-based non-volatile LiM architectures. Due to FeRAM's high reliability, high speed, scalability and CMOS compatibility, it has achieved significant academic and industrial attention \cite{ramaswamy2023nvdram, okuno2023highly}.
  % To leverage FeRAM's potential, Slesazeck et al. \cite{slesazeck20192tnc} proposed a 2T-nC FeRAM cell, suitable for compute-in-memory applications. However, challenges such as destructive read operations and write-back requirements remain.  %ferroelectric memory cell, which is well-suited for compute-in-memory and neuromorphic applications. However, despite its versatility, this design still faces challenges inherent to FeRAM, such as destructive read operations and the need for writeback.
\begin{figure}[t]
    \centering
    \includegraphics[width=0.425\textwidth]{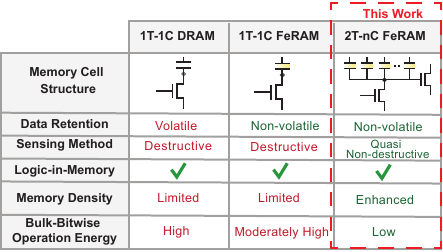}
    \vspace{-3mm}
    \caption{Comparison of 1T-1C DRAM, 1T-1C FeRAM, and 2T-nC FeRAM.}
 \vspace{-18pt}
    \label{fig_1}
\end{figure}
 
% \vspace{-2pt}
Xiao et al. \cite{xiao2023quasi} introduced a quasi-nondestructive readout (QNRO) sensing-based 2T-nC FeRAM for Process-in-Memory (PiM) applications and later proposed a 2T-(n+1)C cell for AND-OR logic functions in memory \cite{10238503}, although it remains complex to program. Existing ferroelectric LiM cells still lack a flexible, configurable approach to universal logic operations without external modifications. To this end, we propose a single-cell in-place logic-in-memory approach using QNRO-enabled 2T-nC FeRAM, requiring no additional peripheral augmentation. This design supports energy-efficient, compact, and high-bandwidth operations with robust reliability. We validate the functionality with fabricated devices and leverage simulations to demonstrate the viability and benefits at an architectural scale.
The major contributions of this paper are:% as follows:
\begin{itemize}

    \item Presents that QNRO sensing inherently supports inverting logic without requiring external circuit modifications. Explores QNRO-enabled LiM in 2T-nC FeRAM unit cell.

    % \item Explores the feasibility of LiM in a QNRO enabled 2T-nC FeRAM unit cell.

    % \item Demonstrates that QNRO sensing enables NOT logic in 2T-nC without external circuit modification.

    %\item \textcolor{red}{We demonstrate that our 2T-nC and higher area density compared to conventional CMOS}

    \item Demonstrates reliable MINORITY function implementation in 2T-nC FeRAM, enabling universal NAND-NOR logic using SPICE simulations and experimental validation. 

     % \item Presents SPICE simulations and experimental verification for implementing reliable MINORITY function associated with universal NAND-NOR logic in 2T-nC FeRAM. We validate functionality with fabricated devices and leverage simulation to analyze more scaled technologies. 
     
     \item Shows that the proposed 2T-nC is compatible with 3D integration, improving storage and computation density.

      \item Evaluates eight real-world data-intensive applications that utilize bulk-bitwise operations, demonstrating significant improvement in energy and performance.

      \item  Validates the thermal viability of the stacked 2T-nC FeRAM, demonstrating its stable and safe operation when placed on a
      compute die.

\end{itemize}

%The rest of the paper is organized as follows. Section II provides background of FE capacitor-based memory devices. Section III details the proposed LiM mechanism and SPICE simulations. Section IV discusses the experimental results. Section V presents workload analysis of bulk-bitwise operations implemented via LiM, followed by conclusion in section VI.

\section{Background}

Process-in-memory (PiM) technologies leverage core memory cells for computation of logic operations, reducing data movement overhead. Figure~\ref{fig_1} presents a comparison of such memory cells, including the conventional 1T-1C DRAM, %alongside FE capacitor-based non-volatile memory (NVM) structures namely
1T-1C FeRAM and 2T-nC FeRAM, both FE capacitor-based non-volatile memories. While 1T-1C DRAM remains the standard,  %memory cell with an optimized structure, 
various bulk-bitwise logic operations, such as bitwise NOT and AND-OR, have been explored in \cite{seshadri2017ambit}. Bitwise NOT uses a Dual-Contact-Cell (DCC) for inversion, while AND-OR relies on Triple-Row-Activation (TRA) to infer output from the majority state. However, DRAM’s destructive reads and volatile storage require frequent copying and refresh.

\begin{figure}[t]
    \centering
    \includegraphics[width=0.48\textwidth]{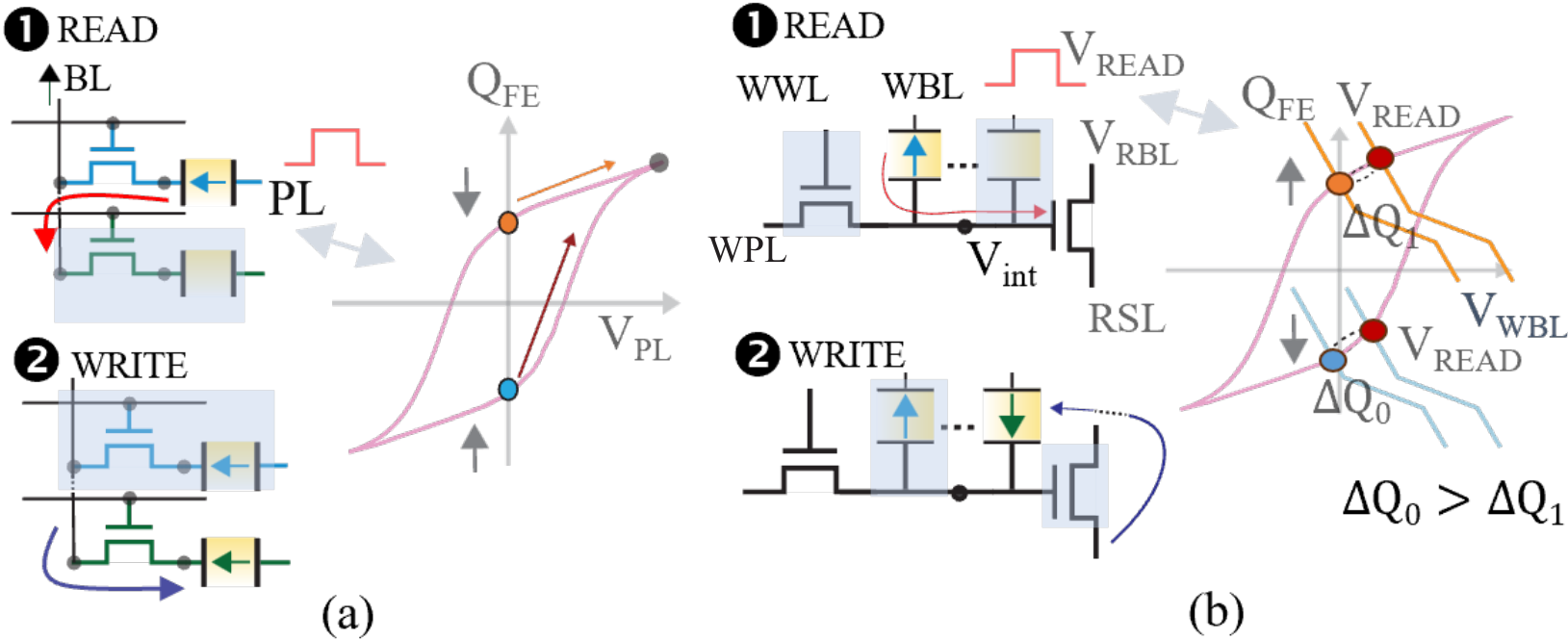}
    \vspace{-0.3cm}
    \caption{(a) Charge sensing non-inverting read and write in 1T-1C FeRAM and (b) QNRO sensing resulting in inverting read and write in 2T-nC FeRAM.}
    \vspace{-16pt}
    \label{fig_2}
    
\end{figure}

In contrast, traditional 1T-1C FeRAM integrates an FE capacitor within a DRAM-like structure \cite{okuno2023highly}. It retains data via nonvolatile ferroelectric polarization, eliminating refresh cycles. When plate line (PL) is activated (Figure~\ref{fig_2}(a)), a large current appears only when the bit line (BL) rises for a state ‘1’ polarization (\ballnumber{1}). The data written to another cell remain unchanged, preserving the source polarization (\ballnumber{2}). However, sensing is still destructive and requires a large write pulse, leading to high energy dissipation.

% In contrast, 1T-1C FeRAM integrates FE capacitor while maintaining DRAM's structure. The FE capacitor retains data through nonvolatile ferroelectric polarization, thus eliminating the need for refresh cycles. The plate line (PL) is activated (figure~\ref{fig_2}(a)), and a large current is detected only when the bit line (BL) rises for the polarization of state ‘1’ (\ballnumber{1}). The data written to another cell remains unchanged and retains the same polarization state as the source (\ballnumber{2}). However, a large write pulse is required for polarization switching, making 1T-1C FeRAM sensing destructive and also dissipating high energy.
% Unlike conventional FeRAM, 1T FeFET supports non-destructive reads, where polarization-dependent charge distributions create distinct threshold voltages. The read voltage, set between these values, avoids disturbing polarization. However, due to FeFETs high write voltages, the endurance is limited to about $10^8$ cycles.

To address the above constraints, the 2T-nC FeRAM architecture enhances performance and endurance by decoupling read/write paths and introducing QNRO. It comprises a write transistor ($T_W$), a read transistor ($T_R$), and multiple ferroelectric (FE) capacitors (Figure~\ref{fig_2}(b)). This configuration improves storage efficiency by reducing transistors per FE capacitor. % compared to 1T-1C. 
The separate paths ensure high write performance and reliability, like 1T-1C FeRAM, while read amplification through $T_R$ enables scalability and read performance similar to 1T FeFET, offering improved read/write control \cite{First_Demo}. During write, $T_W$ directs charge to the target FE capacitor, setting polarization ($P_{FE}$) via complementary voltages on  the write bit line (WBL) and write plate line (WPL). During read, $T_W$ is off; a small read voltage ($V_R$) on WBL enables non-destructive sensing. For positive $P_{FE}$, $V_R$ causes minimal switching ($\Delta Q_1$) (\ballnumber{1}), keeping effective capacitance and internal voltage ($V_{int}$) low, yielding small $T_R$ current. For negative $P_{FE}$, V$_R$ induces greater switching ($\Delta Q_0$), raising capacitance and V$_{int}$, producing higher current through $T_R$.  The current difference is sensed to generate the output and invert the state during transfer (\ballnumber{2}), improving data retention and reliability. Although the switching is significantly lower than conventional FeRAM sensing, it allows multiple reads before $P_{FE}$ changes due to accumulative switching disturb, minimizing write-backs and enhancing endurance ($>10^6$ cycles).

\section{Proposed Design \& Simulation}

\begin{figure*}[t]
    \centering
    \includegraphics[width=0.8\textwidth]{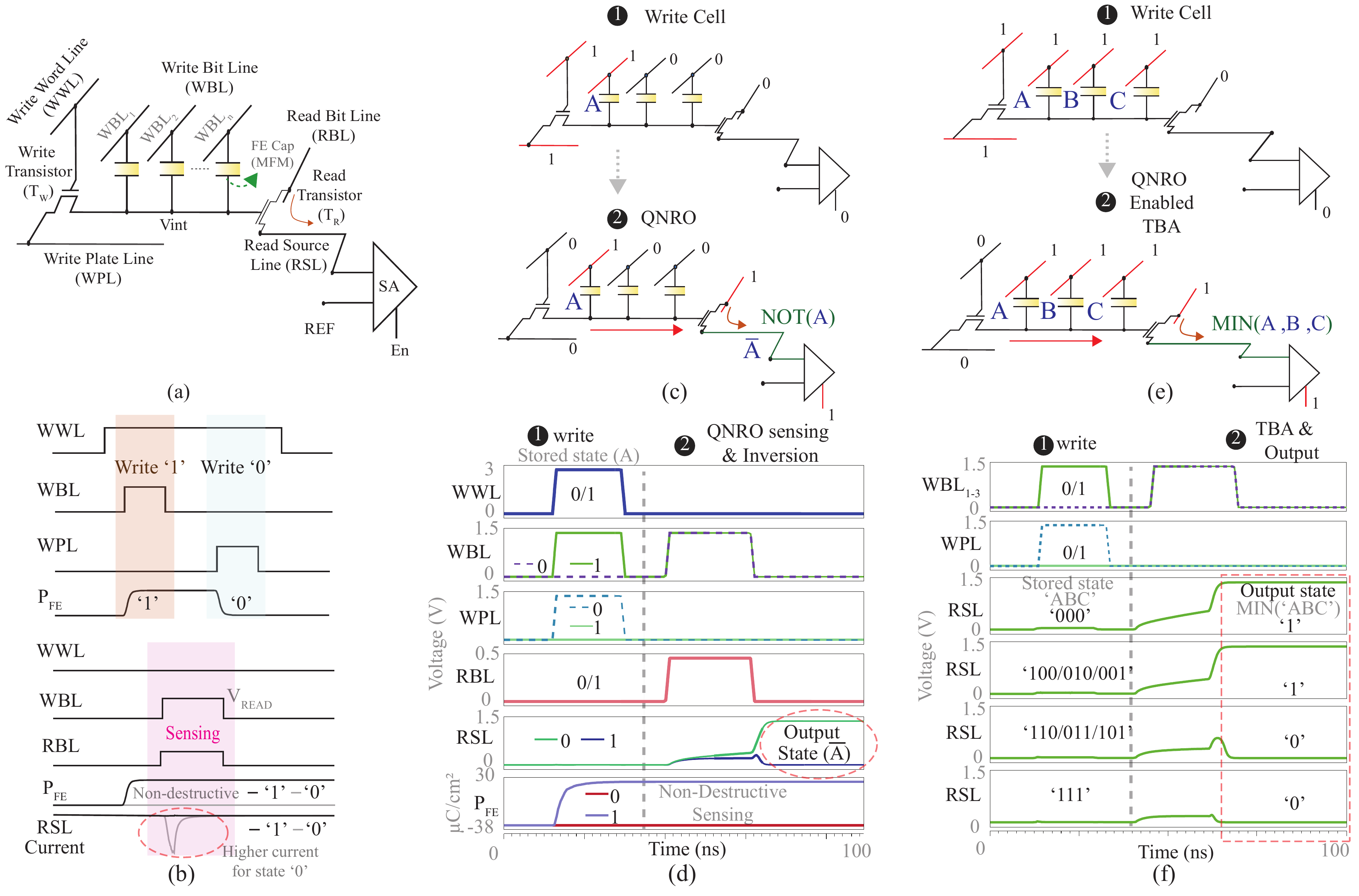}
    \vspace{-0.3cm}
    \caption{2T-nC FeRAM (a) structure (b) writing and sensing mechanism  (c) bitwise-NOT using QNRO (d) SPICE simulation of NOT operation, (e) state transitions for bitwise-NAND-NOR operations, (f) SPICE simulation of bitwise-NAND-NOR operations by performing Triple-Bit-Activation (TBA). }
        \vspace{-16pt}
    \label{fig_2tnc_mech}
\end{figure*}

We extend the multi-cell bitwise logic operations from conventional DRAM \cite{seshadri2017ambit} to non-volatile, QNRO sensing enabled, single cell based operation in 2T-nC FeRAM, validating our approach through circuit simulation.

We utilize Spectre, a SPICE-compatible simulator from Cadence, for netlist based circuit simulations. %\cite{kundert2006designer}.
The NMOS and PMOS transistors were modeled using ASU 45 nm high-performance Predictive Technology Model (PTM) \cite{10.1109/ISQED.2006.91} , and the metal-ferroelectric-metal (MFM) model as FE capacitor following \cite{mfm_source}, calibrated to Micron's NVDRAM cell \cite{ramaswamy2023nvdram} which accurately captures measured ferroelectric behaviors, including device performance scaling, variation, stochastic switching, and domain switching accumulation.

\subsection{Basic Operations in 2T-nC FeRAM}
% The 2T-nC FeRAM cell consists of T$_W$, T$_R$, and multiple MFM capacitors. As illustrated in Figure~\ref{fig_2tnc_mech}(a), this configuration enables a denser arrangement of FE capacitors by reducing the number of transistors per capacitor compared to the 1T-1C structure, thereby improving storage efficiency.

Figure~\ref{fig_2tnc_mech}(a) illustrates a unit 2T-nC cell structure. The write operation begins by raising the WBL of the selected MFM capacitor and activating the write word line (WWL) to enable T$_W$ and connect it to the WPL, as shown in Figure~\ref{fig_2tnc_mech}(b)(top). The polarization state is then programmed to either ‘0’ or ‘1’ by applying controlled voltage pulses to WBL and WPL. This ensures reliable data storage while minimizing unintended disturbances, as illustrated in Figure~\ref{fig_2tnc_mech}(c) (\ballnumber{1}). Multiple capacitors within a single cell can also be written simultaneously in one cycle, as shown in Figure~\ref{fig_2tnc_mech}(e) (\ballnumber{1}).

During read, T$_W$ is turned off. Following the scheme in Figure~\ref{fig_2tnc_mech}(b)(bottom), $V_R$ is applied to the WBL of the target cell, while a small voltage is simultaneously applied to the read bit line (RBL). Data is retrieved by sensing the T$_R$ current at the read source line (RSL), which depends on the ferroelectric capacitor's polarization. A positive $P_{FE}$ (bit '1') causes minimal switching, resulting in low internal voltage ($V_{int}$) and RSL current; conversely, a negative $P_{FE}$ (bit '0') induces greater switching, increasing the RSL current, which is read out using the voltage sensing method \cite{lee2025cross}.

\subsection{NOT Operation in 2T-nC FeRAM}
The execution of NOT operation in 2T-nC FeRAM does not require any external circuit modification such as Dual-Contact-Cell (DCC) in 1T-1C structures \cite{seshadri2017ambit}. Instead, the operation relies on the difference in the T$_R$ current during readout: a high current when reading a stored ‘0’ and a low current when reading a stored ‘1’ during a QNRO. The sense amplifier (SA) detects the current difference and outputs inverted data.

Figure~\ref{fig_2tnc_mech}(c) illustrates the NOT operation in a 2T-nC FeRAM cell. The process begins by writing the initial state (‘0’ or ‘1’) to the selected cell (\ballnumber{1}). During this phase, $T_W$ is activated, while T$_R$ remains off, ensuring that the data is reliably stored. After writing, T$_W$ is deactivated and the stored data is retrieved by enabling T$_R$. As a result, the high T$_R$ current for a stored ‘0’ and the low T$_R$ current for a stored ‘1’ are measured at the RSL. These are compared with a reference level in the SA, which produces an inverted output: a high state (‘1’) for an input state of ‘0’ and a low state (‘0’) for an input state of ‘1’, performing the NOT operation (\ballnumber{2}). 

The SPICE simulation results, shown in figure~\ref{fig_2tnc_mech}(d), further validate this behavior. The simulation demonstrates that after initially writing a ‘0’ or ‘1’ to an MFM capacitor, the sensing process produces logical inversion of the original state. In particular, unlike in 1T-1C FeRAM, the initial state remains fairly intact after readout, preventing unintended data loss.
\vspace{-0.08 cm}
\subsection{Universal Logic Operation in 2T-nC FeRAM}
By leveraging the inverting nature of QNRO, the 2T-nC FeRAM structure can be utilized to perform universal NAND and NOR logic operations through the execution of Triple-Bit-Activation (TBA) in a unit cell. As illustrated in figure~\ref{fig_2tnc_mech}(e), consider three capacitors in a 2T-nC cell storing initial bit states (A, B, C), capable of holding any combination of bits from ‘000’ to ‘111’. When these three bits are simultaneously activated (TBA) by raising the corresponding WBLs and RBL, the resulting output voltage depends on their initial bit states as written in (\ballnumber{1}). This output voltage is then compared to a reference in the SA (\ballnumber{2}), determining the final output state. The resulting output follows the MINORITY (MIN) function, which can be expressed as:
$\text{MIN}(A, B, C) = \overline{C(A+B) + \overline{C} (A \cdot B)}$

 By adjusting the value of the control bit C, TBA enables the implementation of NAND or NOR operations between bits A and B, facilitating efficient in-memory logic computation within the 2T-nC FeRAM architecture.

Figure~\ref{fig_2tnc_mech}(f) presents the SPICE simulation results for the 2T-nC NAND-NOR implementation. The initial states of the cells are set to all combinations between ‘000’ and ‘111’. Upon executing TBA, the resulting current in the RSL is sensed by the SA. The final output state follows the MINORITY of the initial states, realizing NAND-NOR logic in a single cell. 

\begin{figure*}[t]
    \centering
    \includegraphics[width=0.85\textwidth]{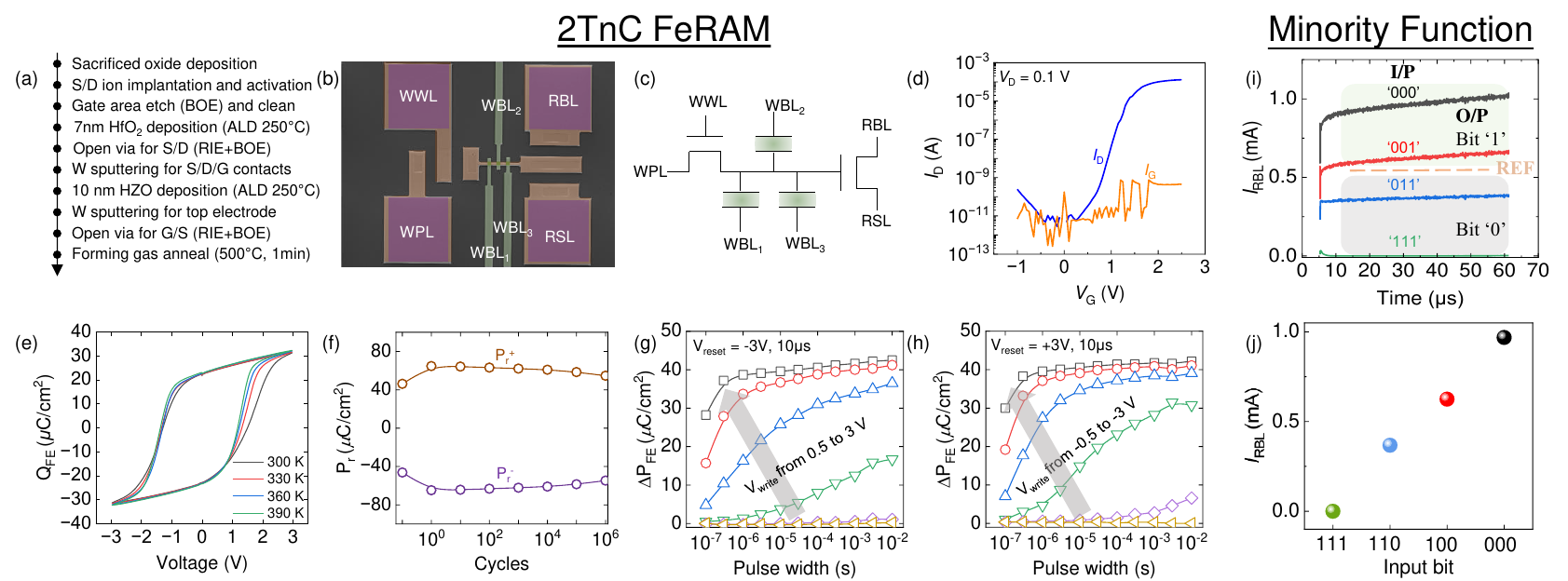}
    \vspace{-0.4cm}
    \caption{(a) Process flow, (b) top-view SEM image, and (c) the schematics of the fabricated 2T-nC FeRAM cell. (d) Transfer curve of the fabricated MOSFET. P-V loop of the fabricated MFM capacitor. (e) P-V loop of the fabricated MFM capacitor at different temperature.(f) Endurance of the MFM capacitor, and corresponding switching dynamics for (g) positive and (h) negative switching. (i) Output current at BL during read. (l) Sensed current at RBL from data “000” to “111” in a 2T-nC FeRAM array. (j) MINORITY function output for different input bits.}
    \vspace{-6mm}
    \label{Experiment}
\end{figure*}

\vspace{-0.1cm}
\section{ Experimental Verification}
The MINORITY function has been experimentally demonstrated in a fabricated 2T-nC FeRAM cell. The fabrication process of the 2T-nC FeRAM cell is shown in figure~\ref{Experiment}(a), with its scanning electron microscope (SEM) top-view image in figure~\ref{Experiment}(b) and schematics in Figure~\ref{Experiment}(c). After fabrication, the individual MFM capacitor and transistor components were characterized. The transfer curve of the transistor, shown in figure~\ref{Experiment}(d), demonstrates an on/off ratio of 10$^7$ and a subthreshold swing (SS) of 110 mV/dec.

The polarization-voltage loop, measured from 300 K to 390 K and spanning voltages from -3 V to 3 V (Figure~\ref{Experiment}(e)), confirms the ferroelectricity of the MFM capacitor, with a remanent polarization ($P_r$) of 22.3 $\mu C$/cm$^2$. The coercive voltage decreases with increasing temperature, while the remanent polarization remains nearly constant.

Endurance testing was performed using multiple $\pm$3 V, 10 $\mu s$  bipolar pulses. The results, shown in figure~\ref{Experiment}(f), reveal that the MFM capacitor can withstand at least 10$^6$ cycles, demonstrating excellent endurance, which is critical for frequent computation. Figure~\ref{Experiment}(g) and (h) show the switching behavior at varying pulse widths and amplitudes for both positive and negative switching, indicating that the MFM can switch with pulse widths under 300 ns at $\pm$3 V.

The MINORITY function was then validated in the 2T-nC configuration. After different data were written to three MFM capacitors, the RBL, or the $T_R$ drain current , was measured. Figure~\ref{Experiment}(i) and ~\ref{Experiment}(j) show the output current at the RBL for different stored data values, from '000' to '111'. The results exhibit an opposite trend, with perfect linearity between the input bit and output current, in contrast to the 1T-1C FeRAM. This demonstrates the ability to compute the MINORITY function when a comparator with a reference voltage is added between the output currents for input bits '001' and '011'. This allows distinguishing '000' and '001' from '011' and '111'.

This experiment validates the operation of FeRAM at relatively high voltages. However, similar operations can be achieved with low-voltage FE capacitors through process optimization \cite{okuno2023highly}. Our measurements indicate that the multi-domain issue does not significantly affect the 2T-nC cell, as the remanent polarization is satisfactory and no severe imprint impact was observed.

\begin{figure*}[]
    \centering
    \includegraphics[width=0.9\textwidth]{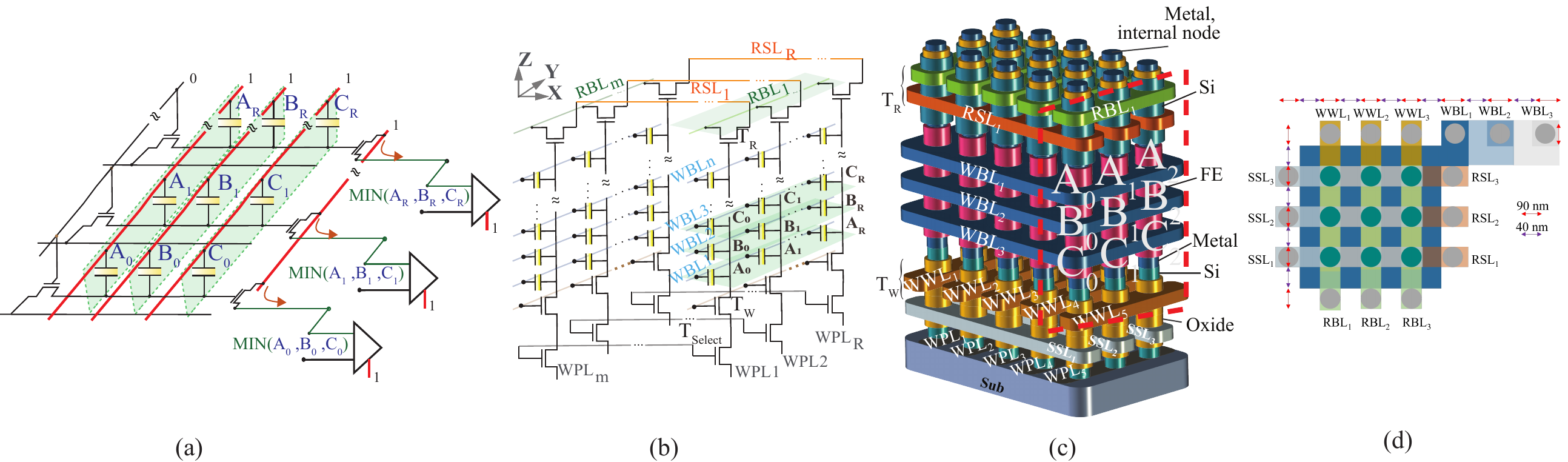}
    \vspace{-0.3cm}
    \caption{(a) Bulk-bitwise operation distribution along a row, (b) vertically stackable 3D array circuit schematic, (c) vertical 3D integration with operand mapping and row activation, and the layout of the (d) 3X3 vertically stacked 2T-3C structure.}
        \vspace{-6mm}
    \label{fig_array_3d_comb}
\end{figure*}
\vspace{-0.2cm}
\section{Bulk-Bitwise Operation \& Vertical 3D integration compatibility} 
\label{bulk}
The in-place logic operations demonstrated at the unit-cell level can be systematically extended to enable row-wise execution across multiple memory cells, facilitating parallel in-memory computation. As shown in~\ref{fig_array_3d_comb}(a), a multi-bit NAND or NOR operation between vectors 
(A\textsubscript0,A\textsubscript1,…,A\textsubscript R) and 
(B\textsubscript0,B\textsubscript1,…,B\textsubscript R) can be mapped by writing each operand bit into the corresponding capacitors of the 2T-nC cells along the row. Simultaneously, control bits 
(C\textsubscript0,C\textsubscript1,…,C\textsubscript R) are configured to define the logic operation.
Once the data and control bits are set, a TBA is applied through the shared WBLs of the selected row by activating the respective RBL. This coordinated activation enables all cells in the row to execute the logic function simultaneously, thus increasing the computational bandwidth.
For row-wise bitwise NOT operations, a similar mechanism is employed. By activating a shared WBL, a NOT operation is executed concurrently across the entire row. This parallelism significantly improves energy efficiency and supports scalable logic-in-memory execution for data-intensive applications.

% \section{Vertical 3D integration compatibility} 

Another key advantage of the 2T-nC structure is its compatibility with vertical 3D integration, enabling significant area savings over traditional planar layouts. As shown in Figure~\ref{fig_array_3d_comb}(b), a vertically stacked 2T-nC FeRAM schematic allows compact integration of transistors and capacitors along a vertical string. Extending the concept from \cite{First_Demo}, Figure~\ref{fig_array_3d_comb}(c) presents a 3D view of a 2T-3C structure, where ferroelectric (FE) capacitors are stacked in the BEOL between the read (T$_R$) and write (T$_W$) transistors. This configuration allows operands to be written directly onto the stacked FE capacitors without incurring extra lateral area. Each vertical string supports bitwise LiM execution, which can be scaled across the row by activating the respective RBL. This significantly increases the compute-per-area density, which can be further enhanced by stacking multiple such layers vertically.

% Another key advantage of the 2T-nC structure is its compatibility with vertical 3D integration, which reduces the area overhead compared to a traditional planar layout. Figure~\ref{fig_array_3d_comb}(b) illustrates a feasible vertically stacked 3D schematic of a 2T-nC FeRAM structure.
% % In contrast to planar distribution, this vertical arrangement optimizes area utilization. 
% Extending from \cite{First_Demo}, figure~\ref{fig_array_3d_comb}(c) provides a 3D perspective of the vertically integrated 2T-3C structure, highlighting how back-end-of-line (BEOL) FE capacitors can be efficiently stacked between the T$_R$ and T$_W$ transistors on a vertical string. Without requiring additional area, the operands are written on the vertically stacked FE capacitors. Each string is capable of performing bitwise LiM, which can be extended along the row by enabling the corresponding RBL. Hence, operation per unit area is increased significantly. Moreover, this benefit can be amplified by stacking multiple layers vertically, further increasing the computational density.

As reported in \cite{28nm_2T1C}, for 28nm technology node, the 2T-1C FeRAM unit cell occupies approximately $30F^2$ area (with $F = 28$nm), where each FE capacitor accounts for $1F^2$. Extending this design for a 2T-3C configuration would result in an estimated area of $\sim90F^2$. In contrast, the vertical stack achieves a compact area of $\sim130 \times 130\,\text{nm}^2$ \cite{First_Demo}, reducing the footprint area per cell  ~4.18 times, thus increasing storage and LiM operation density. 
% As illustrated in figure~\ref{fig_array_3d_comb}(e), the TBA operation per unit area using the vertical stack is ~4.18 times higher than the planar structure, with a higher density of logic operations through multiple layers.

\vspace{-0.1cm}
\section{Workload Evaluation}

% \begin{table*}[!]
%     \centering
%     \scriptsize % Shrinks font size further
%     \caption{Analytical Modeling Assumptions: Energy and Latency Parameters per row (8192B) for DDR4, 1T1C FeRAM, and 2TnC FeRAM}
%     \label{tab:assumptions}
%     \begin{tabular}{lcccccc}
%     \hline
%     \textbf{Technology} & \textbf{ACT Energy (nJ)} & \textbf{ACT Latency (ns)} & \textbf{PRE Energy (nJ)} & \textbf{PRE Latency (ns)} & \textbf{REF Energy (nJ)} & \textbf{REF Latency (ns)} \\ \hline
%     DDR4   & 2.82 & 14.16 & 0.038 & 14.16 & 2.88 & 260.25 \\
%     FeRAM 1T1C  & 2.85 & 14.16 & 0.038 & 14.16 & --    & --    \\
%     FeRAM 2TnC   & 2.63 & 14.16 & 0.039  & 14.16 & --    & --    \\ \hline
%     \end{tabular}
% \end{table*}

\setlength{\tabcolsep}{3pt} % Reduces the column padding (default is 6pt)
\renewcommand{\arraystretch}{0.7} % Reduces row height (default is 1)

\begin{figure}[]
    \centering
    \includegraphics[width=0.45\textwidth]{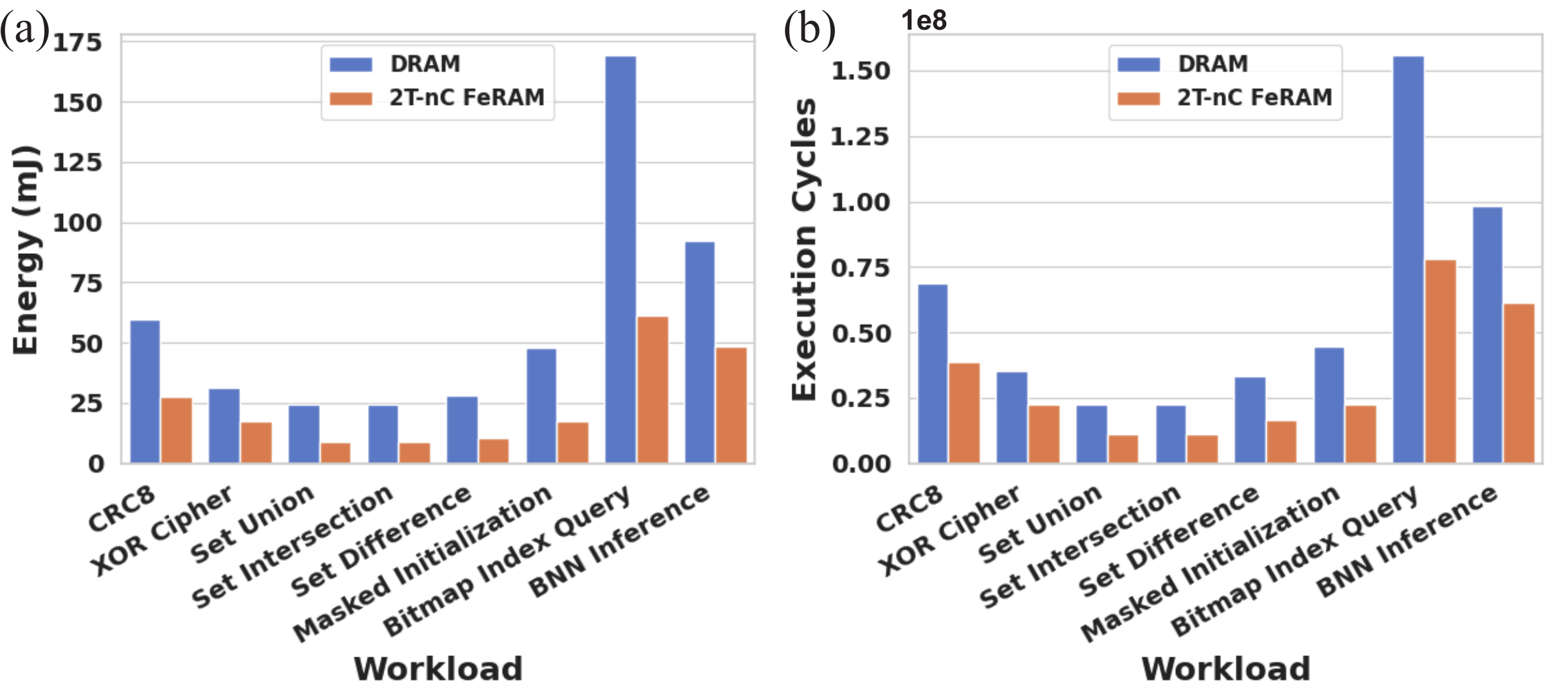}
    \vspace{-12pt} 
    \caption{ (a) Energy consumption and (b) execution cycles for eight real-world, data-intensive workloads executed on DRAM and 2T-nC FeRAM.}
    \label{fig_workload}
    \vspace{-16pt} 
\end{figure}

% We extend the pLUTo simulator \cite{ferreira2022pluto} to model 1T-1C FeRAM, 2T-nC FeRAM, and enhance its DDR4 model by incorporating a refresh cycle analysis. In our setup, each memory technology is simulated with an 8GB capacity, with each row size being 8192 bytes. Figure~\ref{fig_workload} presents a comparative analysis of energy consumption and execution time for eight real-world, data-intensive applications: CRC8 \cite{Warren2013}, XOR Cipher \cite{han1999optical}, Set Union, Set Intersection, Set Difference, Masked Initialization \cite{peleg1996mmx}, Bitmap Index Query \cite{chan1998bitmap}, and BNN Inference \cite{kohut2004boolean}, each with a workload size of 1GB. 
We extend the pLUTo simulator \cite{ferreira2022pluto} to model 2T-nC FeRAM and enhance its DRAM model by incorporating a 64 ms DRAM refresh cycle. For each memory technology, we simulated an 8 GB memory with an 8 KB row size. Figure~\ref{fig_workload} presents a comparative analysis of energy consumption and execution time for eight real-world, data-intensive applications  \cite{seshadri2017ambit}: CRC8, XOR Cipher, Set Union, Set Intersection, Set Difference, Masked Initialization, Bitmap Index Query, and BNN Inference, each with a 1GB workload size. 

% Workloads are mapped to bulk bitwise operations that are executed using row-level primitives. For DDR4 and FeRAM 1T1C, these operations are simulated using an AAP (ACTIVATE-ACTIVATE-PRECHARGE) primitive \cite{seshadri2017ambit}, which leverages the MAJORITY function to perform in-memory AND-OR operations. First, an ACTIVATE command is issued to perform a TRA. Next, a second ACTIVATE command targets the destination row, copying the result via the shared read/write pathway on the bitlines using RowClone, which is enabled by the previous activation. Finally, a PRECHARGE resets the row, making it ready for the next operation.

These workloads are executed as bulk-bitwise operations using row-level primitives. For DRAM, operations are simulated using an AAP (ACTIVATE-ACTIVATE-PRECHARGE) primitive \cite{seshadri2017ambit}, which exploits the MAJORITY function for in-memory AND-OR logic. The first ACTIVATE performs a TRA, the second triggers RowClone \cite{rowclone} to copy data to the destination row via shared bitlines, and the PRECHARGE resets the row for future use.

% For FeRAM 2TnC, bitwise operations are simulated with an ACP (ACTIVATE-COPY-PRECHARGE) primitive using the MINORITY function, which supports universal NAND and NOR logic operations. Initially, an ACTIVATE command executes the TRA. Then, a COPY command writes the RSL data to the destination cell, as the RowClone method is not applicable due to the separate read and write paths. During the copy operation, the data stored in the RSL buffer is transferred to the destination cell via a tri-state buffer, which channels the input from the RSL buffer to the destination cell for writing. Lastly, a PRECHARGE command resets the row for the subsequent operation.

For 2T-nC FeRAM, bitwise operations use an ACP (ACTIVATE-COPY-PRECHARGE) primitive based on the MINORITY function, enabling universal NAND/NOR logic. With $n = 3$, TBA-based logic is executed per cell. ACTIVATE triggers the TBA, COPY transfers RSL data to the destination via tri-state buffer (since RowClone is inapplicable due to separate read/write paths), and PRECHARGE resets the row for the next operation.

For workload simulation model, we use the energy parameters derived based on our cell-level SPICE simulation. The ACTIVATE energy for DRAM and 2T-nC FeRAM is 22.6 nJ and 16.6 nJ per row, respectively. The QNRO mechanism in 2T-nC FeRAM enables this lower energy operation by avoiding the need for a full polarization reversal during read cycles. The PRECHARGE command resets the RSL buffer across all memory cell technologies with an energy of ~0.32 nJ per row. A uniform latency of 1 cycle is assumed for each of the ACTIVATE, PRECHARGE, and COPY operations for both FeRAM and DRAM.

% A uniform 15 ns per cycle is assumed for ACTIVATE and PRECHARGE for both FeRAM and DRAM \cite{ferreira2022pluto}.

% \begin{figure}[]
%     \centering
%     \includegraphics[width=0.45\textwidth]{Figures_SOCC/Mapping_v3.pdf}
%     \vspace{-12pt} 
%     \caption{ (a) Workload mapping, and (b) example of XOR operation breakdown.}
%     \label{fig_workload}
%     \vspace{-10pt} 
% \end{figure}

\begin{figure*}[t]
    \centering
    \includegraphics[width=0.8\textwidth]{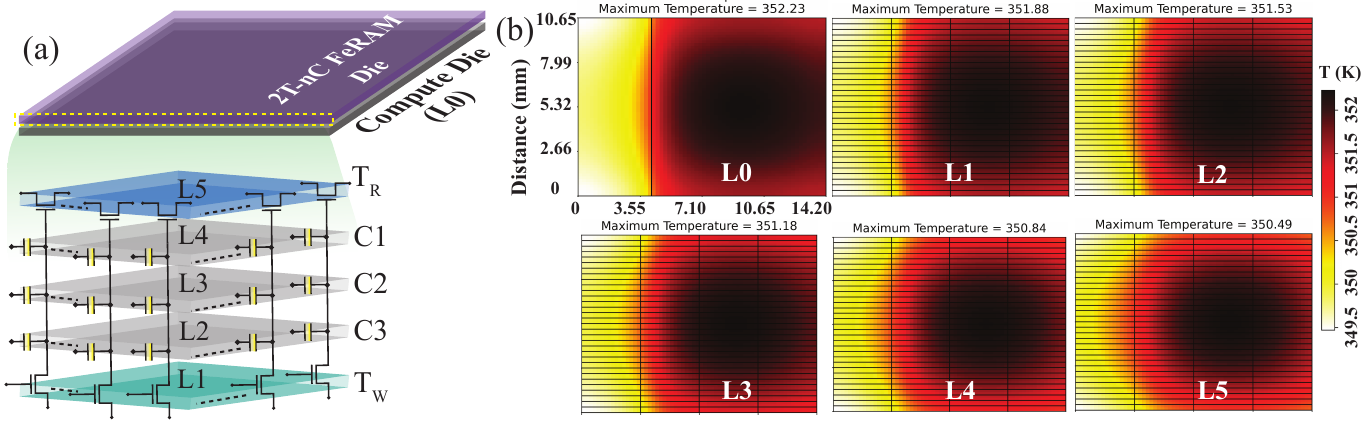}
    \vspace{-0.3cm}
    \caption{ (a) 3D System-on-Chip architecture with a (n+2) 5-layer 2GB verticle 2T-nC FeRAM die stacked on a compute die\cite{tpu}. Memory sub-arrays are shown as individual blocks. (b) Steady-state thermal profile during Bitmap Index Query execution, displaying temperature distribution across the stacks.}
    \label{fig_thermal}
    \vspace{-6mm}
\end{figure*}
\raggedbottom

% In DDR4 and FeRAM 1T-1C designs, destructive reads require that operands be copied to designated rows in order to perform in-memory bitwise operations. These extra copy cycles add to both energy and performance overhead. In contrast, the 2T-nC FeRAM design employs non-destructive reads, eliminating the need for these additional copy cycles and further enhancing energy efficiency and performance.

In DRAM designs, destructive reads require that operands be copied to designated rows in order to perform in-memory bitwise operations. These extra copy cycles add to both energy and performance overhead. In contrast, the 2T-nC FeRAM design employs non-destructive reads, eliminating the need for these additional copy cycles and further enhancing energy efficiency and performance.

% Figure~\ref{fig_workload}(a) shows that FeRAM achieves a 1.13$\times$ reduction in energy consumption compared to DDR4, largely due to eliminating periodic refresh cycles, a significant energy overhead in DDR4. Moreover, the 2T-nC FeRAM design reduces energy consumption even further, by 3$\times$ compared to DDR4. This efficiency is achieved by supporting in-place execution of bulk bitwise operations, which removes the need for separate copy cycles that would otherwise consume extra energy. 

The 2T-nC FeRAM design significantly outperforms DRAM in both energy efficiency and performance, as shown in figure~\ref{fig_workload}. Our proposed 2T-nC memory model achieves a 2.5$\times$ reduction in energy consumption and a 2$\times$ performance improvement. These gains are primarily driven by its ability to execute in-place bulk-bitwise operations. This capability eliminates the costly data copy cycles required in conventional systems, simultaneously increasing performance by reducing execution time and saving energy. The energy and performance benefits are further enhanced by the removal of periodic refresh cycles, a significant overhead inherent to DRAM.

\section{Thermal Analysis}
Thermal constraints pose a major challenge for emerging processing-in-memory (PiM) technologies, particularly when processing elements are integrated at the bank, rank, or sub-array level. To evaluate the thermal reliability of our proposed in-situ 2TnC logic memory, we model a 3-D system-on-chip (SoC) comprising a n+2 layer stack of verticle 2T-nC FeRAM mounted on a compute die. The system utilizes an edge Google TPU with 28 W idle power as a representative compute core \cite{tpu}, and the cell footprint is calculated using the methodology in Section~\ref{bulk} with an additional 50\% overhead for peripheral circuitry, consistent with \cite{28nm_2T1C}. To balance accuracy and computational efficiency in our thermal simulation, we model the memory at the subarray granularity and apply power density values accordingly. Thermal behavior is evaluated using HotSpot \cite{hotspot} under natural convection cooling with a 300K ambient temperature.
During the execution of the bitmap index query workload, the steady-state analyses yield a peak temperature of 351.88 K (Figure \ref{fig_thermal}). The thermal profile is consistent across all evaluated workloads. With a simulated photovoltaic loop for a scaled device, we observed that these operating temperatures preserve the ferroelectric properties of the FeRAM, and stable remanent polarization. This result confirms the thermal viability of the proposed 2T-nC FeRAM for a scaled architecture.

\section{Conclusion}
This work highlights the significant advantages of 2T-nC FeRAM-based LiM over conventional DRAM, particularly in energy efficiency and performance. The implementation of QNRO-enabled universal logic operations in a single 2T-nC FeRAM cell demonstrates its versatility for in-memory bulk computation. Our findings confirm up to 2.5× lower energy consumption and 2× higher performance compared to DRAM, reinforcing FeRAM’s potential as a viable alternative. Additionally, the feasibility of vertical 3D integration further improves its scalability for real-world applications. These results establish FeRAM as a promising candidate for next-generation energy-efficient computing systems.
% This work demonstrates that 1T-1C and 2T-nC FeRAM outperform DRAM in energy-efficient bit-serial logic-in-memory operations. FeRAM eliminates refresh overhead, reducing energy by 1.13× over DDR4, while 2T-nC FeRAM improves performance by 1.87× and further lowers energy by 3× through quasi-nondestructive readout. SPICE simulations and experiments validate key logic functions, and its vertical 3D integration potential makes 2T-nC FeRAM a strong candidate for future logic-in-memory systems.

\section*{Acknowledgment}
% This material is based upon work supported by the center for 3D Ferroelectric Microelectronics and Manufacturing (3DFeM2), an Energy Frontier Research Center funded by the U.S. Department of Energy, Office of Science, Office of Basic Energy Sciences Energy Frontier Research Centers program under Award Number  DE-SC0021118.
The concept development and experiments were supported by the center for 3D Ferroelectric Microelectronics and Manufacturing ($3DFeM^2$),  an Energy Frontier Research Center funded by the U.S. Department of Energy, Office of Science, Office of Basic Energy Sciences, through the Energy Frontier Research Centers program (Award No. DE-SC0021118). The circuit and thermal simulations were supported by NSF Grants 2312886 and 2132918.
%\newpage

%\apptocmd{\thebibliography}{\scriptsize}{}{}
\bibliographystyle{IEEEtran}
\bibliography{ref}

\vspace{12pt}
\balance
\end{document}